\documentclass[aip,cha,amsmath,amssymb,reprint]{revtex4-1}
\usepackage{colortbl}
\usepackage[table]{xcolor}
\usepackage{array}
\usepackage{booktabs}
\usepackage{graphicx}
\usepackage{dcolumn}
\usepackage{bm}

\usepackage{color}

\begin{document}

\title{Generation and cessation of oscillations: Interplay of excitability and dispersal in a class of ecosystem}

\author{Ramesh Arumugam}
\affiliation{Department of Mathematics, Indian Institute of Technology
  Ropar, Rupnagar 140 001, Punjab, India.}

\author{Tanmoy Banerjee}
\email{tbanerjee@phys.buruniv.ac.in}
\affiliation{Chaos and Complex Systems Research Laboratory, Department of Physics, University of Burdwan, Burdwan 713
  104, West Bengal, India.}

\author{Partha Sharathi Dutta}
\thanks{Corresponding author\\}
\email{parthasharathi@iitrpr.ac.in}
\affiliation{Department of Mathematics, Indian Institute of Technology
  Ropar, Rupnagar 140 001, Punjab, India.}

\received{:to be included by reviewer}
\date{\today}

\begin{abstract}
 
We investigate the complex spatiotemporal dynamics of an ecological
network with species dispersal mediated via a mean-field coupling.
The local dynamics of the network are governed by the
Truscott--Brindley model, which is an important ecological model
showing excitability.  Our results focus on the interplay of
excitability and dispersal by always considering that the individual
nodes are in their (excitable) steady states. In contrast to the
previous studies, we not only observe the dispersal induced generation
of oscillation but we also report two distinct mechanisms of cessation
of oscillations, namely amplitude and oscillation death.  We show
that, the dispersal between the nodes influences the intrinsic
dynamics of the system resulting multiple oscillatory dynamics such as
period-1 and period-2 limit cycles.  We also show the existence of
multi-cluster states which has much relevance and importance in
ecology.

\end{abstract}


\maketitle

\begin{quotation}

Species dispersal among connected habitats often identifies the
complex spatial dynamics of ecological system and significantly
increases the persistence of ecological communities for longer
time. Various dynamical models have been used to describe the effect
of dispersal in connected habitats. As far as ecological models are
concerned, sometimes there exists slow-fast time scales with very
interesting dynamics. For example, in aquatic ecosystem, plankton
bloom is a result of sudden changes in environmental fluctuations that
makes plankton ecosystem as excitable media. Take this into account,
the effect of dispersal in slow-fast dynamical ecological system is
analyzed qualitatively using mean-field assumption as an external
force. In a homogeneous environmental set up, the coupled slow-fast
system shows multiple characteristics of sensitivity in synchronized
oscillations for different initial density.

\end{quotation}

\section{Introduction}
\label{sec:intro}

Excitability is one of the interesting features of slow-fast dynamical
systems that is characterized by the fact that a small perturbation in
the input leads to a large excursion in phase space before coming to
the rest state \citep{Lon00}.  Notably, an excitable medium possesses
stable equilibria which exhibits qualitatively different behavior
(large excursion in the phase space) according to the character of an
external perturbation \citep{Fran13,Olla13}.  In most of the physical
and biological systems, excitation arises with various dynamical
aspects.  In particular, in neuronal systems, two main types of
excitability are defined, namely type-I and type-II excitability.  The
type-I excitability is characterized by the appearance of a stable
limit cycle with arbitrarily low frequency via a global bifurcation
\citep{Lind04,Kea12}.  On the other hand, the type-II excitability
yields zero-amplitude and finite period spikes through the
supercritical Hopf bifurcation \citep{Lai03}.

Although, the notion of excitability has been well studied in the
context of neuronal systems \citep{Izh00}, it remains less explored in
the field of ecology, where excitability plays an important role in
maintaining species diversity, e.g., in aquatic ecosystems
\citep{TrBr94}.  As far as the excitable ecological systems are
concerned, external perturbation arises naturally in the form of
demographic rates \citep{Ste85}, environmental
fluctuations\citep{ChHu97, BjGr01}, seasonal variation\citep{Wa02},
and even migration of populations \citep{Amar08, BlHu99}.  Therefore,
it is of natural interest to explore the role of excitability in
ecological systems.

Furthermore, like neurons, ecological systems are also rarely isolated
\citep{TiKa98}.  The dispersal of species through an external force
often connects the fragmented habitats \citep{ShCh02} which
subsequently promotes the synchronized oscillations \citep{BaDu15}.
In other words, the connectivity of habitats through migration
(coupling) enhances the relationship between synchrony and stability.
Hence, it is of broad interest to examine the collective behaviors of
interacting excitable ecological units. In literature, a large number
of studies have been devoted to explore the collective behaviors of
excitable units in biology, such as, neurons, genetic oscillators,
beta cells in islets of Langerhans, etc \citep{Izh00}. In all these
studies, interaction takes place in a {\it microscopic scale}, e.g.,
through the sharing of membrane voltage or diffusion of
ions. Depending upon the underlying mechanisms of the individual nodes
and their organization, these interactions are governed by coupling
topology. Therefore, the natural question to ask is how the similar
types of coupling functions affect the collective behaviors of an
excitable system in a {\it macroscopic scale} such as an {\it
  ecological} network with excitable units? This study is relevant
since in both the biological and ecological networks the types of
interactions are quite identical: For example diffusion or quorum
sensing mechanism through ions in biology is equivalent to the
dispersal or weighted mean-field dispersal of species density in an
ecological system. Therefore, in the present study we try to reveal
the following important questions: Does the generation of oscillation
rely on the characteristics of excitable system or type of coupling we
used in?  What is the effect of dispersal on the dynamics of excitable
ecological systems?  What are the new dynamical features involved in
this coupled excitable systems?

To address these questions, we emphasize on the excitable system's
features by considering an ecological system, namely the
``Truscott--Brindley model'' with species spatial movement.  In the
context of ecological systems, the Truscott--Brindley model determines
the excitability due to fluctuating weather conditions with
demographic and environmental noise as perturbations
\citep{Mor09}. Concerning the external forces of an ecological system,
here we use mean-field coupled Truscott--Brindley model as a
consumer--resource model in which migration of populations takes place
among the selected habitats and preserves the fundamental
characteristics of an excitable system.  Generally, this mean--field
assumption is used as a diffusive coupling in physical
\citep{BaBi13,BaGh14,BaGh14a}, biological \citep{strz01} as well as in
ecological systems \citep{BaDu15} to quantify the average
distribution.  In contrast to excitable oscillations, in this paper we
show that the oscillatory behaviour of individual patches are
suppressed through two distinct mechanisms, namely amplitude death
(AD) and oscillation death (OD).  In general, the oscillation
quenching mechanisms such as AD and OD play important roles to
suppress the oscillations in most of the physical, chemical and
biological oscillators \citep{KoVo13}. Basically, AD is the mechanism
by which two or more interacting oscillators arrive at a common
homogeneous steady state, whereas in OD oscillators
populate different branches of stable inhomogeneous steady states
which are created by symmetry breaking in the network
\citep{KoVo13,KoVo13a}. The OD state is particularly important
from biological point of view as it induces inhomogeneity in an
otherwise homogeneous network that has relevance in biology, e.g., in
synthetic genetic oscillator \cite{kosepl,koschaos} and cellular
differentiation \cite{cell}. It should be noted that although the
occurrence of AD has been reported earlier in excitable ecological
oscillators under conjugate coupling \cite{KaRaFe14} but this is the
first time we reveal the occurrence of oscillation death and other
interesting behavior, such as, multi-cluster oscillation death. With
the presence of active and passive dispersal in homogeneous and
heterogeneous habitats we further reveal the multiple characteristics
of excitable system such as generation of oscillation and transition
from period-1 to period-2 limit cycle with certain threshold value in
the parameter.

The outline of this paper is as follows.  First, in
Sec.~\ref{sec:model}, we explain the uncoupled and the coupled TB
model with dispersal in only consumer population.  In
Sec.~\ref{sec:res}, with variations in physiological and environmental
parameters, various qualitative behaviours of dispersal effect with
appearance and disappearance of oscillations are described for
identical and non-identical patches.  Following that, bidirectional
coupling is taken into account, various dynamical consequences of
dispersal and excitability are illustrated in Subsec.~\ref{sec:asym}.
Finally, to check the robustness of coupled excitable system, a
network of few patches is analyzed and thereafter we have the
discussion and concluding remarks in Sec.~\ref{sec:dis}.

\section{Mathematical Model}
\label{sec:model}

To study the consumer-resource interactions with the excitable (i.e.,
slow-fast) local dynamics in a patchy habitat (i.e., spatially
extended population), we consider the Truscott--Brindley model in each
of the patches.  The Truscott--Brindley model exhibits type-II
excitability, and is known to mimic plankton blooms in ecosystems
\cite{TrBr94}.

\subsection{Non-dimensionalized Truscott--Brindley Model}

\begin{figure}
\centering \includegraphics[width=0.49\textwidth]{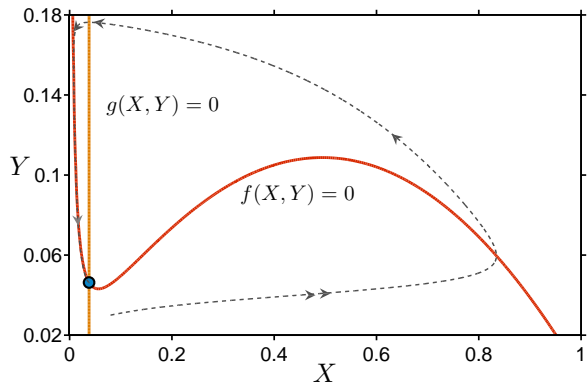}
\caption{(Color online) Nullclines of the resource $(X)$ and the
  consumer $(Y)$ populations in the uncoupled Truscott--Brindley model
  (\ref{eq:1}) are shown here for fixed parameters $\gamma=0.05$,
  $\omega=0.34$, $\nu=0.053$ and $\beta=0.43$. The red curve
  represents the resource equation (\ref{eq:nd1}), whereas the orange
  line represents the consumer equation (\ref{eq:nd2}).  The solid
  circle represents the equilibrium point.  The dashed arrow marked
  curve shows how an initial condition approaches the equilibrium
  point.}
\label{f:null}
\end{figure}

We start with the dimensionless form of the Truscott--Brindley model
\citep{TrBr94,Ed01}.  The dynamics of the resource ($X$) and the
consumer ($Y$) with their associated interactions are given by the
below equations:
\begin{subequations}\label{eq:1}
\begin{align}
 \frac{dX}{dt} & =f(X,Y)= \beta X(1-X)-Y
 \frac{X^2}{X^2+\nu^2} ,\label{eq:nd1} \\ \frac{dY}{dt}
 & =g(X,Y) =\gamma\left(\frac{X^2}{X^2+\nu^2}-
 \omega\right)Y,\label{eq:nd2}
\end{align}
\end{subequations}
where $\beta$ is the maximum growth rate of the resource ($X$), $\nu$
is half saturation constant of the consumer ($Y$) which governs how
quickly maximum predation rate is attained as density of the resource
increases, $\gamma$ is the maximum growth rate of the consumer ($Y$)
or the conversion efficiency rate of the ingested resource due to
predation and further $\omega$ represents the consumer's mortality
rate. The growth of the resource ($X$) is characterized by the
logistic growth function and grazing by the consumer ($Y$) is
represented by the Holling type-III functional response
\citep{Mur_book}.  The model (\ref{eq:1}) has equilibrium point and
oscillatory states for different parametric set up.  To exhibit
slow-fast dynamics the value of $\nu$ must stay in the parameter
range: $0 < \nu < \frac{1}{3\sqrt{3}} = 0.1924$.  This parametric
range ensures that the nullcline $f(X,Y)=0$ has two turning points
(one local maxima and one local minima) those are instrumental for the
system to exhibit slow-fast dynamics (see Fig.~\ref{f:null}). Another
parameter which governs the qualitative behaviour of the model
(\ref{eq:1}) is $\omega$.  Since the position of the nullcline
$g(X,Y)=0$ is determined by the value of $\omega$. The other two
parameters $\gamma$ and $\beta$ do not have much influence on the
qualitative behavior of the Truscott--Brindley model \cite{TrBr94}.
An exemplary parameter values for an equilibrium point are:
$\beta=0.43$, $\nu=0.053$, $\gamma=0.05$ and $\omega=0.34$.  To
identify the collective dynamics of coupled excitable system, we use
this parameter values so that each uncoupled system (\ref{eq:1}) has
only fixed point state.  However, to have oscillations in the
uncoupled model (\ref{eq:1}), the value of $\omega$ needs to be
changed accordingly.

\subsection{Coupled Truscott--Brindley  Model}
In nature, it is commonly understood that the species diversity in
spatially fragmented habitats can be preserved by the movement of
populations from nearby patches. Such movement or dispersal of
populations between spatially separated patches is potentially
important for the survival and persistence of the community
\citep{Han_book}.  Without dispersal, the consumer--resource dynamics
in a single patch is represented by the uncoupled Truscott--Brindley
model Eq.~(\ref{eq:1}).  Here species spatial movement among $N$
number of patches is taken into account, firstly we couple only the
consumer populations ($Y$) in each patch using the mean-field
dispersion \citep{BaDu15}. Therefore, the coupled model due to species
movement is given by:
\begin{subequations}\label{eq:2}
\begin{align}
 \frac{dX_i}{dt} &= \beta X_i(1-X_i)- Y_i \frac{X_i^2}{X_i^2+\nu^2}
 ~,\; \\ \frac{dY_i}{dt} &= \gamma\left(\frac{X_i^2}{X_i^2+\nu^2}-
 \omega\right)Y_i+\epsilon\left(Q\overline{Y}-Y_i\right),
\end{align}
\end{subequations}
where $i=1,2,\hdots,N$ and $\overline{Y}=\frac{1}{N}\sum
\limits_{i=1}^{N} Y_i$.  The parameter $\epsilon$ represents the
coupling strength (or the dispersal rate) of the consumer ($Y$) and
$Q$ represents the mean-field density of the consumer which quantifies
the average distribution of density in the patches.  In fact, this
mean field density ($Q$) determines the qualitative and the
quantitative behavior of migrated consumer populations among the
selected patches \cite{BaDu15, RaDuBa15}.  Each $i$-th patch has two
distinct dynamical features.  One is populations local dynamics within
a patch where population density is quantified due to interaction
between the resource and the consumer.  Another one is dispersal
dynamics due to mean-field coupling or exchange of individuals in
consumer populations ($Y$) between the patches.  Depending on species
density in each patch, the consumer movement can be either emigration
or immigration in their respective patch.


\section{Results}
\label{sec:res}

We analyze this excitable system starting with nullclines of the
uncoupled dimensionless model (\ref{eq:1}).  In Fig.~\ref{f:null}, we
show the nullclines of the consumer and the resource.  Intersection of
this two nullclines is an equilibrium point of the considered model
(\ref{eq:1}).  Also the qualitative behavior of phase space can be
easily identified using these nullclines.  The dashed curve in
Fig.~\ref{f:null} exhibits how a trajectory in $(X,Y)$ phase plane
approaches the equilibrium point following two different time scales
(fast along the $X$ axis and slow along the $Y$ axis).  Further,
linear stability analysis of the coupled system (\ref{eq:2}) is
performed wherever possible, otherwise numerical bifurcation analysis
is carried out using XPPAUT \citep{xpp} package by adopting a suitable
method for stiff systems.

\begin{figure*}
\centering \includegraphics[width=0.99\textwidth]{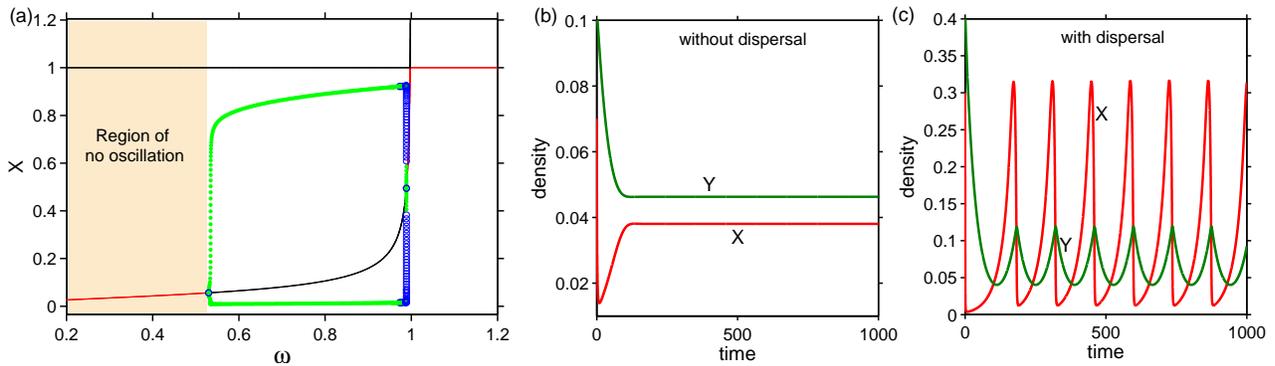}
\caption{(Color online) (a) One parameter bifurcation diagram for
  varying $\omega$ of the uncoupled model (\ref{eq:1}).  Here shaded
  region represents the occurrence of steady state.  Green and blue
  circles represent stable and unstable limit cycles respectively,
  whereas red and black curves represent the stable and unstable
  steady states.  (b) Time series of both resource ($X$) and consumer
  ($Y$) of the uncoupled model (\ref{eq:1}) for $\omega =0.34$.  (c)
  For $\omega=0.34$, time series of both the resource ($X$) and
  consumer ($Y$) in the presence of dispersal when the dispersal rate
  $\epsilon=0.2$ and the mean-field density $Q=0.95$.  Other parameter
  values in (a)-(c) are $\gamma=0.05$, $\nu=0.053$ and
  $\beta=0.43$.}
\label{f:rhy}
\end{figure*}

\subsection{Dynamics of the uncoupled system}

The small value of $\nu$ ($=0.053$) ensures that the system has a
slow-fast dynamics.  For low mortality rate of the consumer (i.e.,
$\omega=0.34$) in the uncoupled model (\ref{eq:1}), no oscillation
occurs and populations exist only in the excitable steady state.  In
Fig.~\ref{f:rhy}(a), we have shown a one-parameter bifurcation diagram
for varying the mortality rate ($\omega$) of the consumer using the
uncoupled Truscott--Brindley model.  From Fig.~\ref{f:rhy}(a), it is
clear that oscillation starts beyond a certain mortality rate, say the
Hopf bifurcation point $\omega_H$ (here $\omega_H=0.528$) in the
uncoupled model (\ref{eq:1}).  In the shaded region, there exists no
oscillation other than excitable steady states.  We show temporal
dynamics of the model (\ref{eq:1}) in Fig.~\ref{f:rhy}(b) that indeed
shows the occurrence of steady state after a brief transient episode.

\subsection{\label{ID} Dynamics of the coupled system: Identical patches }

\begin{figure}
\centering \includegraphics[width=0.47\textwidth]{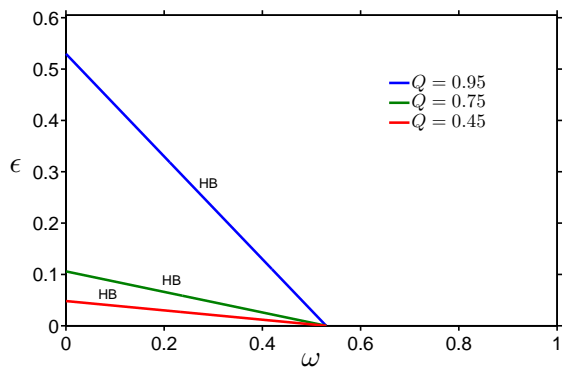}
 \caption{(Color online) Two-parameter bifurcation diagram for
   different mean-field density $Q$: $Q=0.45$, $Q=0.75$, and $Q=0.95$;
   $\omega-\epsilon$ space is shown where the Hopf bifurcation (HB)
   curve separating two different regions of equilibrium point and
   oscillation.  HB curve represents the point where the oscillation
   starts due to the coupling, below HB curve is the region where only
   steady states occur and above HB curve is the region of both
   appearance and disappearance of oscillations. Other parameter
   values are $\gamma=0.05$, $\nu=0.053$ and $\beta=0.43$.}
\label{f:ew}
\end{figure}

\begin{figure*}
\centering \includegraphics[width=0.89\textwidth]{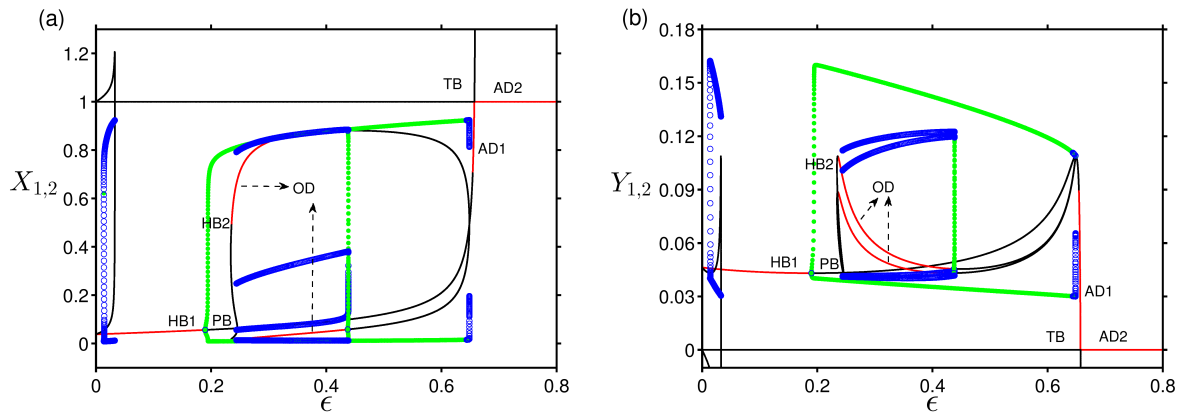}
\caption{(Color online) Dispersal induced rhythm and creation of AD
  and OD; individual patches are in the excitable steady state
  ($\omega=0.34$): (a) One parameter bifurcation diagram of the
  resource ($X$) for varying coupling strength ($\epsilon$).  Here AD,
  OD, HB, PB and TB represent that amplitude death, oscillation death,
  Hopf bifurcation, pitchfork bifurcation and transcritical
  bifurcation, respectively. (b) One parameter bifurcation diagram of
  the consumer ($Y$) for varying coupling strength ($\epsilon$).
  Other fixed parameters are $\gamma=0.05$, $\nu=0.053$, $\beta=0.43$
  and $Q=0.95$.  Here green and blue circles represent the stable and
  the unstable limit cycles, respectively, whereas red and black
  curves represent the stable and the unstable steady states,
  respectively.  }
\label{f:ID_a}
\end{figure*}

\subsubsection{Patches are in excitable steady state: Effects of dispersion}

We consider that the patches are in excitable steady states and
examine the effect of dispersion on the coupled dynamics. Let us start
with an exemplary scenario of generation of rhythm from the steady
state that is induced by the dispersal between the patches: We take
$\omega=0.34$ which is less than $\omega_H$ and uncoupled dynamics are
in the excitable steady state (c.f. Fig.~\ref{f:rhy}(b)). In the
presence of dispersal ($\epsilon=0.2$) and mean-field density
($Q=0.95$) (see Eq.~(\ref{eq:2})), both the consumer and the resource
populations start to oscillate (see Fig.~\ref{f:rhy}(c)).  Thus in
presence of dispersal, the coupled system (\ref{eq:2}) exhibits
rhythmic behavior, where as its uncoupled unit (\ref{eq:1}) is at a
rest state.

Moreover, to explore the generation of oscillation from the excitable
steady state we find the relationship between the coupling parameters
and the local dynamics (governed by $\omega$) using a two parameter
($\omega$--$\epsilon$) bifurcation diagram for three different $Q$
values (Fig.~\ref{f:ew}); the zone below the Hopf bifurcation (HB)
curve represents steady state, which yields oscillation through the
Hopf bifurcation.  From Fig.~\ref{f:ew} it is observed that for
$\omega>\omega_H$ individual patches are in oscillating zone
irrespective of dispersion. But for $\omega<\omega_H$, HB lines become
steeper from low to high mean field density ($Q$).  Note that, for low
value of mean-field density ($Q=0.45$), even for low dispersal rate
($\epsilon$), oscillation occurs (see Fig.~\ref{f:ew}).

Next, we consider two identical patches at their respective steady
states and investigate the appearance of oscillations as well as
oscillation's transition to coupling induced steady states (both
homogeneous and inhomogeneous) through the oscillation quenching
mechanisms such as AD and OD in the coupled model (\ref{eq:2}).  Here,
identical in the sense that local dynamics of consumer and resource
are same for all the patches.  In Fig.~\ref{f:ID_a}, we depict a one
parameter bifurcation diagram with variations in the dispersal rate
($\epsilon$).  Here, each fixed $\epsilon$ and $Q$ represent a local
habitat setup, whereas varying $\epsilon$ or $Q$ represents the
environmental fluctuations in the local habitat.  The qualitative
dynamics of the coupled system is identified in this bifurcation
diagram for varying dispersal rate.  In Fig.~\ref{f:ID_a}(a), steady
states of the resource ($X$) are given for varying $\epsilon$ with
fixed $Q=0.95$.  Other fixed parameters are $\gamma=0.05$,
$\omega=0.34$, $\nu=0.053$, and $\beta=0.43$.  It is important to note
here that initially we start with a stable steady state instead of
oscillatory state in each patch of the uncoupled system.  With an
increase in the parameter $\epsilon$, a steady state is transformed
into oscillatory state at $\epsilon_{HB1} \approx 0.1898$.  Further,
OD is created by symmetry breaking of the steady state through a
pitchfork bifurcation (PB) at $\epsilon_{PB} \approx 0.2456$, whereas
AD is created through a transcritical bifurcation (TB) at
$\epsilon_{TB} \approx 0.6572$.  After the pitchfork bifurcation, OD
creates inhomogeneous steady states through the Hopf bifurcation (HB2)
at $\epsilon_{HB2} \approx 0.2355$ (shown in Figs.~\ref{f:ID_a}(a)).
The consumer ($Y$) dynamics for varying the dispersal rate $\epsilon$
are also shown in Fig.~\ref{f:ID_a}(b).  Notice that, a very small
change in $\epsilon$ may lead to a critical transition from AD1 to
AD2, where suddenly the consumer goes to extinction, i.e., $Y_{1,2}=0$
(see Fig.~\ref{f:ID_a}(b)).  Here AD1 represents the non-zero density
of both the consumer and the resource whereas AD2 is non-zero density
of only the resource and the consumer is extinct from the community.
This is due to the fact that at the TB point (i.e., $\epsilon_{TB}
\approx 0.6572$ where $X_i=1$) the resource attains its maximum
carrying capacity \cite{Mur_book}. Note that as stable limit cycles
share the phase space with the OD state, thus, depending on the
initial conditions, we have either stable oscillations or stable
steady states.  In general, if we vary the dispersal rate $\epsilon$
with changing the mean-field density $Q$, we find similar dynamics,
except now the values of bifurcation points change accordingly.

\begin{figure}
\centering \includegraphics[width=0.495\textwidth]{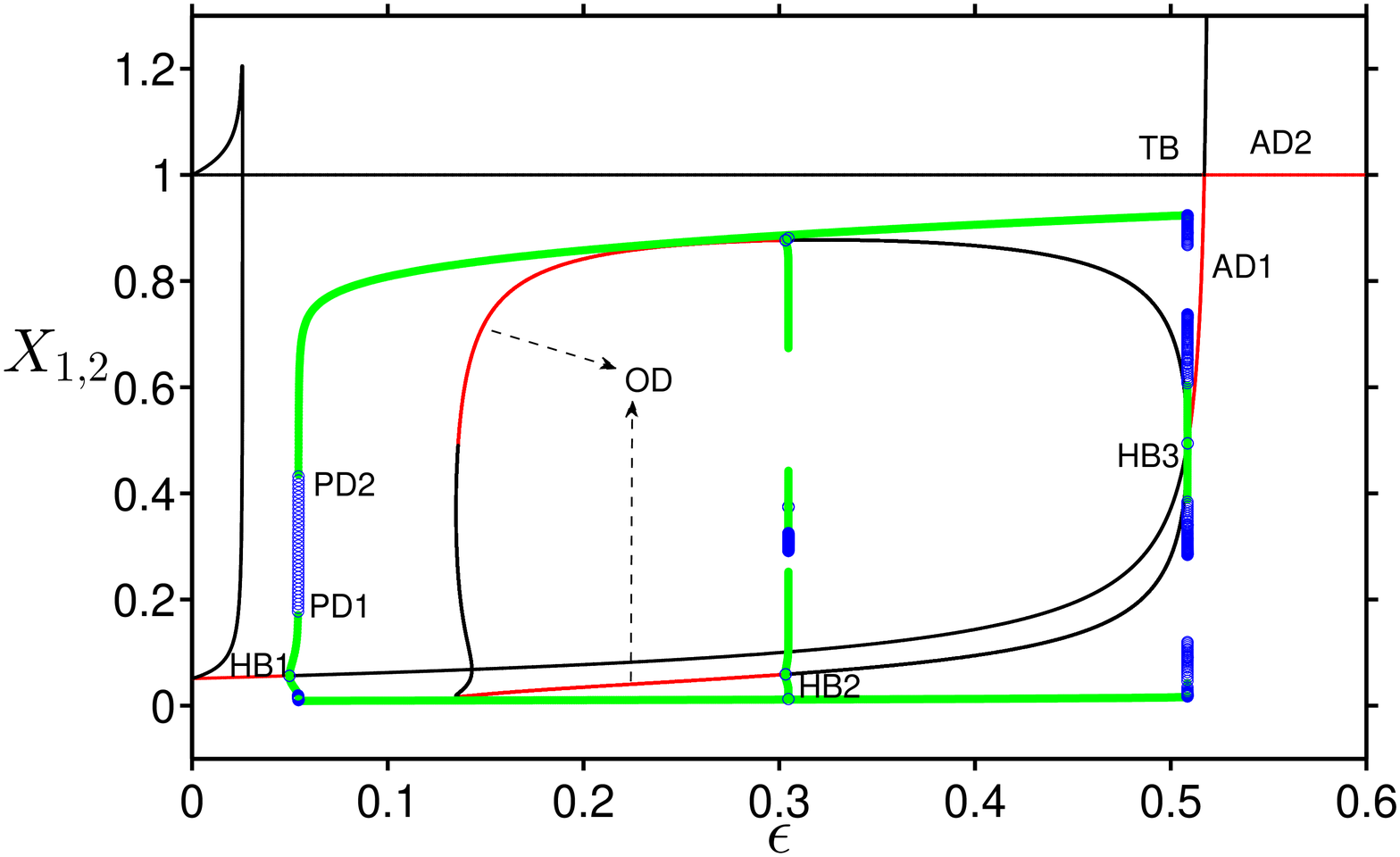}
\caption{(Color online) One parameter bifurcation diagram for varying
  the coupling strength ($\epsilon$) with the fixed mean-field density
  ($Q=0.95$).  Here PD represents the period-doubling bifurcation.
  Other parameters are fixed at $\beta=0.43,~\nu=0.053,~\gamma=0.05$
  and $\omega=0.48$.}
\label{f:ex2}
\end{figure}

\begin{figure*}
\centering \includegraphics[width=0.85\textwidth]{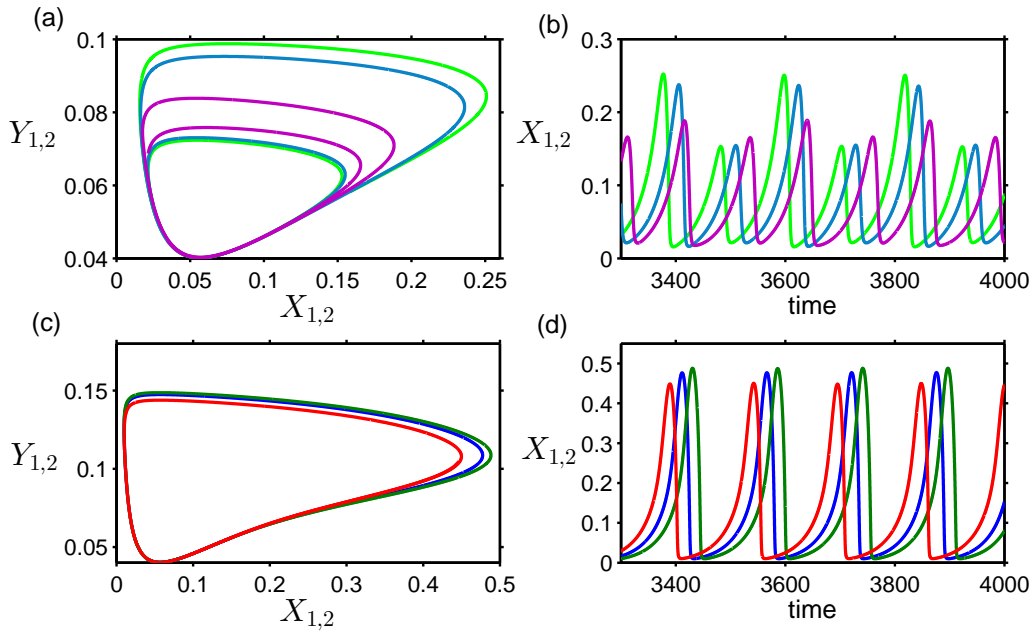}
\caption{(Color online) Phase portrait and temporal dynamics
  exhibiting excitability for fixed $\epsilon=0.0545$ and $Q=0.95$:
  (a) Period-2 limit cycles, where three different initial conditions
  are chosen around the unstable period-1 limit cycles (i.e., blue
  circles between PD1 and PD2 in the Fig.~\ref{f:ex2}(a)).  (b)
  Corresponding time series of period-2 limit cycles in (a).  (c)
  Period-1 limit cycle, where three different initial conditions are
  chosen around the stable limit cycles shown above PD2 in
  Fig.~\ref{f:ex2}(a).  (d) Corresponding time series of period-1
  limit cycles in (c).  Other parameters are same as in
  Fig.~\ref{f:ex2}.}
\label{f:ex3}
\end{figure*}

As $\omega$ is shifted towards the edge of the oscillation, i.e.,
$\omega_H$, we find qualitative changes in the bifurcation
scenarios. Now, with increasing dispersal rate $\epsilon$,
not only we get a transition from steady state to periodic
oscillation, but additionally we observe higher periodic
oscillations through the period doubling bifurcation of limit
cycle. We take $\omega=0.48$ (remember $\omega_H=0.528$) and computed
one parameter bifurcation diagram (Fig.~\ref{f:ex2}).
Importantly, focus the region where the dispersal rate ($\epsilon$) is
between $0.04$ and $0.06$ in Fig.~\ref{f:ex2} and at this region, the
coupled system shows multiple characteristics in a small variation of
initial density and dispersal rate. First, period-1 limit cycle is
created via a Hopf bifurcation (HB1) at $\epsilon_{HB1} \approx
0.04981$.  Then the stable period-1 limit cycle becomes unstable and
there is a creation of a stable period-2 limit cycle via a
period-doubling bifurcation of limit cycle (PD1) at $\epsilon_{PD1}
\approx 0.05417$.  Further, there is a transition from the stable
period-2 limit cycle to a stable period-1 limit cycle via a reverse
period-doubling bifurcation of limit cycle (PD2) at
$\epsilon_{PD2}\approx 0.05442$.  Subsequently, there exists similar
oscillatory characteristics in HB2 at $\epsilon_{HB2} \approx 0.3032$
with OD.

The phase space and time series for a particular choice of the
parameters $\epsilon=0.054$ (i.e., in between HB1 and PD1) are shown
in Figs.~\ref{f:ex3}(a)-(d).  For a small perturbation in initial
conditions, we have distinct time period of oscillation of stable
period-$1$ and period-$2$ limit cycles.  For this parametric set up,
the coupled system shows neutral oscillations ( i.e. center ) since we
have different amplitude and time period for each initial
condition. Further, there is some threshold in initial condition leads
to period-1 limit cycle. Instead of center in the dynamics of
oscillation, above a certain threshold in the dispersal rate lead to
stable limit cycle.  Figs.~\ref{f:ex3}(a) and \ref{f:ex3}(b) show the
phase space and time series of period-$2$ limit cycles, whereas
Figs.~\ref{f:ex3}(c) and \ref{f:ex3}(d) show the phase space and time
series of excitable period-$1$ limit cycles with different time period
of oscillation.  In other words, initial conditions for
Fig.~\ref{f:ex3}(a) are chosen between the period doubling
bifurcations PD1 and PD2 mentioned in the Fig.~\ref{f:ex2}.  As the
Truscott--Brindley model mimics the plankton algal bloom in aquatic
ecosystems, the existence of period-2 cycles resembles the multiyear
cycles in aquatic ecosystems \citep{Klli12}.

From the above results, we have seen that (see Fig.~\ref{f:ex2}) a
stable oscillation is always created from an excitable steady state
through a supercritical Hopf bifurcation confirming the fact that our
present ecosystem belongs to the type-II excitable system.

\subsubsection{Stability Analysis}

Considering the perfect synchrony of the coupled two-patch ecosystem
(\ref{eq:2}), we calculate bifurcation curves/points by using the
fixed points of the coupled system wherever possible.
Eqs.~(\ref{eq:2}) has $(0,0,0,0)$ and $(1,0,1,0)$ as trivial fixed
points, whereas $(X^*,Y^*,X^*,Y^*)$ is a nontrivial fixed point, where
\begin{subequations}
\begin{align*}
 X^{*}&=\sqrt{\frac{\nu^{2} (\epsilon - Q \epsilon + \gamma \omega)
   }{\gamma+ (Q-1)\epsilon - \gamma \omega}}, ~~\mbox{and}\\ Y^{*}&=
 \frac{-\beta \gamma \nu^2 }{(\gamma+ (Q-1)\epsilon - \gamma \omega)}
 \\ ~~~&\;\;\;~ + \frac{\beta \gamma \nu}{\sqrt{(\epsilon - Q \epsilon
     + \gamma \omega)(\gamma+ (Q-1)\epsilon - \gamma \omega)}}~~.
 \end{align*}
\end{subequations}
The Jacobian $\mathcal{J}_{1}$ of the system (\ref{eq:2}) at
$(1,0,1,0)$ is given by:
\[
\mathcal{J}_{1}=
  \begin{bmatrix}
    j_{11}& j_{12} &:&0&0\ \\
     0 & j_{22} &:& 0&j_{24} \\
      \cdots & \cdots &  & \cdots & \cdots \\
      0&0&:&  j_{11} & j_{12}\\
       0& j_{24} &:& 0&j_{22}\ 
 \end{bmatrix},
\]
where,
\begin{eqnarray*}
 j_{11}&=&-\beta,~ j_{12} =-\frac{1}{1+\nu^2},~j_{24}= \frac{\epsilon
   Q}{2},~~ \mbox{and}\\ j_{22}&=&-1+\frac{\epsilon
   Q}{2}+\gamma\left(\frac{1}{1+\nu^2}-\omega\right) ~ .
 \end{eqnarray*}
The eigenvalues of the Jacobian $\mathcal{J}_1|_{(1,0,1,0)}$ are:
\begin{subequations}
\begin{align}
\lambda_{1,2}&=j_{11} =-\beta, ~\mbox{and}  \label{Eig1a}\\ 
\lambda_{3,4}&= j_{22}\mp j_{24} ~. \label{Eig1b}
\end{align}
\end{subequations}

Solving the eigenvalues $\lambda_{3,4}$ (given by (\ref{Eig1b})) for
$\epsilon$ or $\gamma$, we get two bifurcation curves.  In particular,
solving $j_{22}+j_{24}$ for $\epsilon$ gives the transcritical
bifurcation (TB) curve of Fig.~\ref{f:ex2}, where:
\begin{eqnarray*}
 \epsilon_{TB}&=&
 \frac{\gamma(\omega-1+\nu^{2}\omega)}{(Q-1)(1+\nu^2)} ~.
\end{eqnarray*}
For the nontrivial fixed point $(X^{*}, Y^{*}, X^{*}, Y^{*})$,
the Jacobian matrix $\mathcal{J}_{2}$ is given by:
\[
\mathcal{J}_{2}=
  \begin{bmatrix}
    j_{11}^{+}& j_{12}^{+} &:&0&0\ \\
 j_{21}^{+} & j_{22}^{+}  &:  & 0&-j_{22}^{+} \\
 \cdots & \cdots &  & \cdots & \cdots \\
 0&0&:& j_{11}^{+} & j_{12}^{+}\\
 0&-j_{22}^{+}&: & j_{21}^{+}&j_{22}^{+}\
 \end{bmatrix},
\]
where,  $ j_{12}^+ = \frac{(Q-1)\epsilon}{\gamma}-\omega,~\; j_{22}^{+}
= -\frac{Q \epsilon}{2}$,


\begin{widetext}
\begin{eqnarray*}
j_{11}^{+}&=&\frac{2\beta\epsilon(Q-1) \left(\sqrt{\nu^2 (\epsilon-
    Q\epsilon + \gamma\omega)}-\sqrt{\gamma + (Q-1)\epsilon -
    \gamma\omega}\;~\right)} {\gamma \sqrt{\gamma + (Q-1)\epsilon -
    \gamma\omega}}\\
~~~&~& +\frac{\beta \gamma\left(-\sqrt{ \gamma + (Q-1)\epsilon- \gamma\omega} +
  2\omega\left(\sqrt{\gamma + (Q-1)\epsilon- \gamma\omega} - \sqrt{\nu^2
    (\epsilon -Q\epsilon+ \gamma\omega)}~\right)~\right)}{\gamma \sqrt{\gamma +
    (Q-1)\epsilon - \gamma\omega}},~\mbox{and}
\\ j_{21}^{+} &= &-2
\beta\left(\epsilon-Q\epsilon+\gamma(\omega-1) +
\sqrt{\gamma+(Q-1)\epsilon-\gamma\omega}\sqrt{\nu^2(\epsilon-Q\epsilon+\gamma\omega)}\right).
\end{eqnarray*}
\end{widetext}
The eigenvalues of the Jacobian $\mathcal{J}_2|_{(X^*,Y^*,X^*,Y^*)}$
are
\begin{subequations}
\begin{align}
\lambda^+_{1,2} &= \frac{1}{2}\left(j^+_{11}\mp
\sqrt{(j^+_{11})^2+j^+_{12}
  j^+_{21}}\;\right),~\mbox{and} \label{Eig2a}\\ \lambda^+_{3,4} &=\frac{1}{2}
\left(j^+_{11}+2j^+_{22}\mp \sqrt{(j^+_{11}-2j^+_{22})^2+4j^+_{12}
  j^+_{21}}\;\right)~. \label{Eig2b}
\end{align}
\end{subequations}

Solving the real part of the eigenvalues $\lambda_{3,4}^{+}$ in
Eq.~(\ref{Eig2b}) for $\epsilon$ with other parameters fixed, we get
two Hopf bifurcation (HB1 and HB3) curves of Fig.~\ref{f:ex2}.  Due to
cumbersome mathematical expressions, we don't mention those
bifurcation curves here.

\begin{figure}
\centering \includegraphics[width=0.49\textwidth]{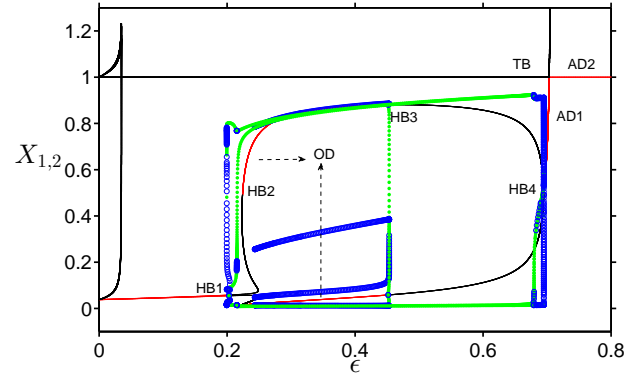}
\caption{(Color online) One parameter bifurcation diagram for
  non-identical patches with variations in $\epsilon$: Resource
  population densities of $X_{1,2}$ are shown along $y$-axis.  The
  limit cycles (green and blue circles) in vertical directions
  represent the excitation which has a different period for distinct
  initial conditions.  Here $\gamma=0.05$ in patch-1 and
  $\gamma_2=0.057$ in patch-2, and the other parameters are same as in
  Fig.~\ref{f:ID_a}.}
\label{f:asym}
\end{figure}

\subsection{Dynamics of the coupled system: Non-identical local dynamics}

In real spatial ecosystems, interacting patches are, in general,
non-identical \citep{Hol_book}. For example, spatial and environmental
heterogeneity due to weather fluctuations make the fragmented habitat
heterogeneous \citep{OpWa04}.  Here we assume heterogeneous patches by
considering distinct conversion efficiency ($\gamma$) of the consumer
population in each patch and look for the transition from the
excitable steady state to oscillation and also to oscillation
suppression mechanisms (AD and OD).  For mismatch in species local
dynamics, we set $\gamma=0.05$ in patch-1 and $\gamma=0.057$ in
patch-2.  Remaining parameters are identical in both the patches.

\begin{figure*}
\centering \includegraphics[width=0.9\textwidth]{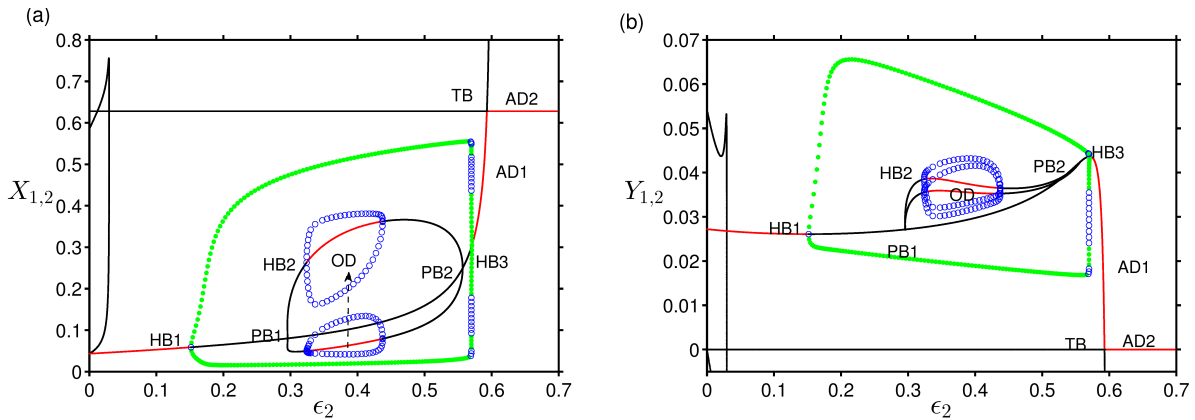}
\caption{(Color online) Asymmetric coupling: One parameter bifurcation
  diagram of: (a) Resource ($X_i$) for varying the coupling strength
  $\epsilon_2$ and (b) consumer ($Y_i$) for varying the coupling
  strength $\epsilon_2$.  Other fixed parameters are
  $\beta=0.43,~\nu=0.053,~\gamma=0.05$, $\omega=0.4$,
  $\epsilon_1=0.2,~ Q_1=0.2$, and $Q_2=0.95$.}
\label{f:bc}
\end{figure*}

Slow-fast oscillation occurs here also, but it takes place with
combination of both stable and unstable limit cycles.  Thus, depending
upon the initial condition the system (\ref{eq:2}) will either
converge to a stable limit cycle or to an unstable limit cycle.
Figure~\ref{f:asym} represents a one parameter bifurcation diagram
with variations in $\epsilon$.  Here OD is created due to the mismatch
in species local dynamics. Apart from the OD and AD states, we have
excitable oscillation with vertically distributed green circles and
blue circles in the mentioned HB points of Fig.~\ref{f:asym}.
Moreover, wherever Hopf bifurcation occurs (i.e., HB1, HB2, HB3 and
HB4), we have slow-fast oscillations in the coupled non-identical
patches.

\subsection{Coupling in both resource and consumer}
\label{sec:asym}
Long term persistence of ecosystem functioning and community structure
involve the collective dynamics of species distribution and
subsequently, to shape the biodiversity, organisms movement plays
important role \citep{Jel13}.  As every organism depends on the other
for their resource in most of the terrestrial and aquatic ecosystems,
while dispersal of the consumer takes place for favorable conditions,
passive dispersal happens in the resource also (i.e., instead of
moving, it's being moved either directly or indirectly).  Moreover,
the complexity of food web dynamics and it's associated interactions
involve the dispersal of all the species presence in the ecosystem
\citep{Amar08, McC05}.  So the collective excitable dynamics can be
strengthened if we consider the coupling in both the variables $X_i$
and $Y_i$.  Instead of only the consumer movement to check the
excitability in the Truscott--Brindley model (\ref{eq:1}), we consider
the more natural dispersal (i.e., when both species are coupled with
mean-field assumption) by setting distinct dispersal rate and
mean-field density in consumer--resource populations.  The coupled TB
model now becomes:
\begin{subequations}\label{eq:bc}
\begin{align}
 \frac{dX_i}{dt} &= \beta X_i(1-X_i)- Y_i
 \frac{X_i^2}{X_i^2+\nu^2}+\epsilon_1\left(Q_1\overline{X}-X_i\right),
 \\ \frac{dY_i}{dt} &= \gamma\left(\frac{X_i^2}{X_i^2+\nu^2}-
 \omega\right)Y_i+\epsilon_2\left(Q_2\overline{Y}-Y_i\right),
\end{align}
\end{subequations}
where $i=1,2,\hdots,N$, $\overline{X}=\frac{1}{N}\sum
\limits_{i=1}^{N} X_i$ and $\overline{Y}=\frac{1}{N}\sum
\limits_{i=1}^{N} Y_i$. Here $\epsilon_1$ and $\epsilon_2$ represent
the dispersal rate of the resource $X_i$ and the consumer $Y_i$,
respectively, whereas mean-field density of $X_i$ and $Y_i$ are $Q_1$
and $Q_2$, respectively.

\subsubsection{Asymmetric coupling between two patches}

Recently, trait-based approaches have been used to predict various
consequences of ecological communities \citep{AcBr15,MeBr14}.  In
ecosystems, characteristics of each community differs as per
environmental conditions due to spatiotemporal heterogeneity in the
habitat by their nature.  So it is natural that the dispersal rate at
the consumer level can differ from the dispersal rate at the resource
level.  As symmetric coupling is a spacial case of the more general
situation of asymmetric coupling, hence, in spatial parameters, we use
asymmetric coupling in the dispersal rate and the mean-field density,
i.e., $\epsilon_1\neq \epsilon_2$ and $Q_1\neq Q_2$ \cite{RaDuBa15}.
We check the coupled system dynamics for varying dispersal rate of
consumer ($\epsilon_2$) where other spatial parameters are fixed at
$\epsilon_1=0.2,~Q_1=0.2$ and $Q_2=0.95$ with the identical local
dynamics in each patch.

\begin{figure*}
\centering \includegraphics[width=0.75\textwidth]{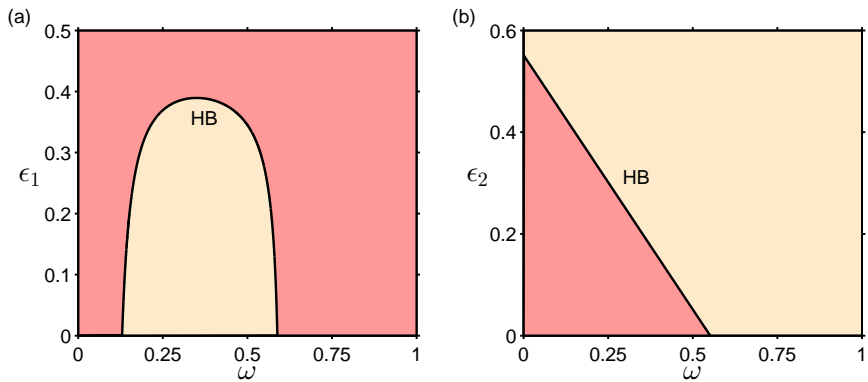}
\caption{(Color online) Generation of Oscillation : (a) Oscillations
  due to asymmetric coupling for the fixed parameters
  $\epsilon_2=0.4$, $Q_1=0.2$ and $Q_2=0.95$, (b) oscillations are
  created at HB due to asymmetric coupling with fixed parameters
  $\epsilon_1=0.2$, $Q_1=0.2$ and $Q_2=0.95$. Here, the red shaded
  region represents the occurrence of only fixed steady states, HB
  curve represents the point where the oscillation starts due to the
  coupling and light color shaded region after HB curve is the region
  of both appearance and disappearance of oscillations.  Other fixed
  parameters are $\beta=0.43$, $\nu=0.053$, $\omega=0.34$ and
  $\gamma=0.05$. }
\label{f:we_bc}
\end{figure*}

Here AD and OD occurs along with slow-fast oscillations.
Figs.~\ref{f:bc}(a) and \ref{f:bc}(b) show both appearance and
disappearance of oscillations using one-parameter bifurcation diagram
of the resource ($X$) and the consumer ($Y$) for varying dispersal
rate of the consumer $(\epsilon_2)$.  Although we start with steady
state in the uncoupled model, in the coupled system, creation of
oscillations occurs at HB1 ($\epsilon_{HB1} \approx 0.1516$). Further,
here also OD is created by symmetry breaking of steady state through
pitchfork bifurcation (PB1) at $\epsilon_{PB1} \approx 0.2957$.
Further, after HB3 (at $\epsilon_{HB3}\approx 0.5706$), oscillations
are suppressed and give raise to homogeneous steady state (AD1).
Interestingly, a small increase in dispersal rate leads to the
transition from AD1 to AD2 through transcritical bifurcation (TB) at
$\epsilon_{TB} \approx 0.5929$.

Figure~\ref{f:we_bc} shows two-parameter bifurcation diagrams those
identify the occurrence oscillations for each dispersal rate through
HB. For asymmetric coupling, this is shown in Figs.~\ref{f:we_bc}(a)
and \ref{f:we_bc}(b) with $\omega$--$\epsilon_1$ and
$\omega$--$\epsilon_2$ planes respectively.  In fact, here steady
states only occur in the red shaded region and oscillation
starts at HB point whereas light color shaded region is for
coexistence of oscillation along with AD and OD states.  Moreover,
even though we have started with fixed point in the uncoupled system,
in Fig.~\ref{f:we_bc}(a), oscillation arises in light color shaded
region even at $\epsilon_1=0$ due to the dispersal of species from the
other patch (i.e., $\epsilon_2 \neq 0$).


\subsubsection{ A network with more than two patches}

As spatial movement of species takes place in a large number of
patches in natural ecosystems, we consider a network (\ref{eq:bc})
with $N=32$ patches where dispersal takes place in both the
populations.  Moreover, both spatial and environmental heterogeneity
are taken into account and we analyze the mean-field coupled network
in two distinct cases.

\begin{figure*}
\centering \includegraphics[width=0.97\textwidth]{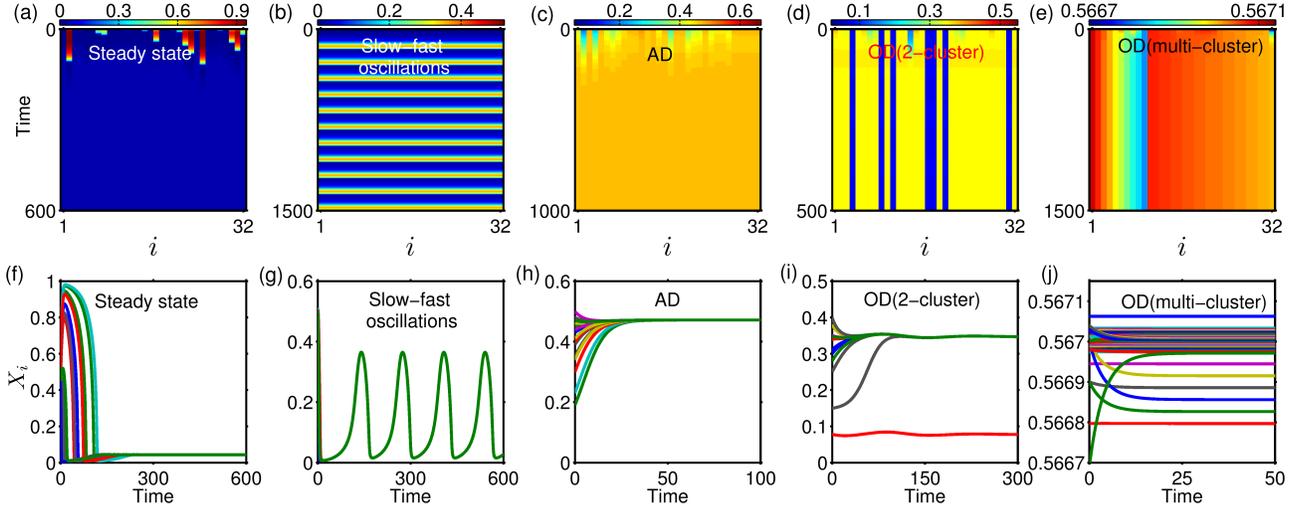}
\caption{(Color online) Dynamics of a network with $N=32$ nodes (i.e.,
  patches): (a) Steady states of 32-patches when uncoupled for the
  fixed parameters $\epsilon_1=\epsilon_2=0$, (b) Slow-fast
  oscillations through asymmetric coupling for the fixed parameters
  $\epsilon_1=0.2$ and $\epsilon_2=0.1647$, (c) AD due to asymmetric
  coupling for the fixed parameters $\epsilon_1=0.2$ and
  $\epsilon_2=0.5875$, (d) OD for asymmetric coupling with identical
  local dynamics in 32 patches for the parameters $\epsilon_1=0.2$ and
  $\epsilon_2=0.5145$, (e) OD with multi-cluster due to mismatch in
  local dynamics with parameters $\epsilon_1=0.2$ and
  $\epsilon_2=0.608$.  In (f)-(j) are the time series corresponding to
  the spatial dynamics shown in (a)-(e) respectively.  Other fixed
  parameters are $Q_1=0.2$ and $Q_2=0.95$, $\beta=0.43$, $\nu=0.053$,
  $\gamma=0.05$ and $\omega=0.4$.}
\label{f:ns1}
\end{figure*}

{\it Case--I}: First, we consider asymmetry in the spatial parameters
(i.e., $\epsilon_i$ and $Q_i$) with identical local dynamics, i.e., in
Eqs.~(\ref{eq:bc}) we consider $\epsilon_1\neq\epsilon_2$ and $Q_1\neq
Q_2$.  In this case, the creation of oscillation, AD and OD are still
possible. The spatiotemporal dynamics of oscillation generation is
shown in Figs.~\ref{f:ns1}(b) and \ref{f:ns1}(g), whereas the dynamics
of it's uncoupled version is shown in Figs.~\ref{f:ns1}(a) and
\ref{f:ns1}(f).  When there is no dispersal, each patch is in a steady
state, then dispersal makes the network of connected patches to show
the synchronized oscillation which is further suppressed to
AD/OD. Here, the spatiotemporal dynamics of AD is shown in
Figs.~\ref{f:ns1}(c) and \ref{f:ns1}(h) for $\epsilon_1=0.2$,
$\epsilon_2=0.5875$, $Q_1=0.2$ and $Q_2=0.95$, whereas spatiotemporal
dynamics of OD is shown in Figs.~\ref{f:ns1}(d) and \ref{f:ns1}(i) for
the parameters $\epsilon_1=0.2$, $\epsilon_2=0.5145$, $Q_1=0.2$ and
$Q_2=0.95$.

{\it Case--II}: Next, we use mismatch in species local dynamics
together with asymmetric coupling. In particular, we set mismatch in
consumer's conversion efficiency ($\gamma$) for all $N=32$ patches.
We choose the mismatch in each patch in the following way:
$\gamma_{i}=\left(1+\frac{i}{100}\right)\gamma_0$, where
$i=1,2,\hdots,32$ and $\gamma_0=0.05$. Due to mismatch in species
local dynamics, inhomogeneous steady states are created and thus forms
a multi-clustered OD state which is shown in Figs.~\ref{f:ns1}(e) and
\ref{f:ns1}(j) for the parameters $\epsilon_1=0.2$,
$\epsilon_2=0.608$, $Q_1=0.2$ and $Q_2=0.95$.  In multi-clustered OD
state, populations populate in different steady states and the
position of those steady states may change depending upon the choice
of $\gamma_{i}$ and the initial conditions.

Thus, Fig.~\ref{f:ns1} generalizes the results of this excitable
system by depicting the spatiotemporal dynamics starting from steady
states of the uncoupled system, creation of oscillations and it's
transition to oscillation suppression such as AD and OD with 2-cluster
and multi-clustered states when both populations are coupled.
However, if dispersal takes place in only one species, the qualitative
dynamics remains same. Also the qualitative nature of the network
remains same if one considers more number of patches.

\section{Discussion and conclusion}
\label{sec:dis}

In summary, in this paper we have explored the emergent behaviors of
an excitable ecological network local dynamics where the dynamics in
each node are governed by the Truscott--Brindley model.  The
connections between the nodes are governed by the mean-field
dispersion.  We have emphasized on the interplay of excitability and
dispersal by always considering that the individual patches are in
their (excitable) steady states. Unlike previous studies,
\citep{sin11,Das10} we have not only observed the generation of
oscillation (or so called rhythmogenesis) but we have also reported
two distinct mechanisms of cessation of oscillations, namely amplitude
and oscillation death.  We have analyzed various dynamical aspects of
dispersion using spatial and environmental heterogeneity. At first,
considering dispersal only in the consumer population and also the
local habitat's interconnection, the generation of oscillations is
identified in the coupled system from their respective steady states.
The mean-field dispersion assumption used here, resembling as a
globally coupled excitable system, potentially determines the
qualitative behaviors of slow-fast dynamical systems along with
excitation.  Typically, the dispersal between the patches influences
the intrinsic dynamics of the system resulting multiple oscillatory
dynamics such as period-1, period-2 limit cycles, center and stable
limit cycle.  Moreover, the excitable dynamics due to variation of
only initial conditions in the interacting habitats show synchronized
stable oscillations with distinct time period of oscillations.  It is
important to note that the population movement changes the intrinsic
dynamics of the uncoupled system and promotes generation of
oscillation in the coupled system.  While the species dispersion
between the patches generate the oscillations by exhibiting the
type-II excitability in the coupled Truscott--Brindley model, on
contrary, the same coupling feature is used to suppress the
oscillations.  In fact, the multiple oscillation suppression states,
namely AD and OD highlight the transition of oscillation to stable
homogeneous and inhomogeneous steady states respectively in a
homogeneous patchy habitat.  Essentially, the mean-field coupling
constitutes both appearance and disappearance of oscillations
alongside excitable dynamics both in identical and non-identical
patches.

On the other hand, the consumer and the resource dispersal are taken
into account, we have analyzed the consequences of excitability
through asymmetric coupling along with a network of globally connected
patches.  Our findings indicate that the combined effect of species
dispersion for varying species local dynamical parameters as well as
spatial parameters also determines the excitability.  In addition to
that, oscillation quenching states (AD and OD) are determined in both
homogeneous and heterogeneous habitats.  Overall, in all the results,
we have started with steady states in an uncoupled Truscott--Brindley
model and then moved to oscillatory state through mean-field coupling
and further transitioned to stable steady states (i.e., AD and OD).
This dynamical phenomena of back and forth behavior of steady state to
oscillations is valid for a large number of connected nodes (i.e.,
patch) also. So the species dispersal can self-assemble the ecological
communities among the fragmented habitats which prevents the complete
extinction of local species.  However, instead of starting from steady
state in the uncoupled Truscott--Brindley model, if each individual
patch is in oscillatory state, then the oscillation quenching
mechanisms of the coupled system is still valid.  In fact, the
excitation with AD and OD is also valid in that case.

While the long term dynamics of plankton organisms is based on the
dynamics of Truscott--Brindley model, the dispersion phenomena has
significant effect in structuring the community in ecosystems.
However, the environmental fluctuations as a stochastic effect have an
impact in climatic dynamics due to the seasonal cycles variation of
planktonic organisms. In line with the reason that the tininess of
plankton organisms, mean-field description is a suitable one to study
the species density in plankton ecosystems \citep{Olla13, Mor12}.
Considering the dispersal as like fluctuations, our coupled
Truscott--Brindley model localizes the populations regionally through
mean-field description and show the excitability with various
behaviors of limit cycles due to perturbation.  Hence our findings may
be helpful for the regulation and the restoration of the stable as
well as the oscillatory populations using the back and forth behavior
of coupling features.  Overall, the mean-field description enables
both rhythmic and steady state behaviors of an ecological system.
Further the complexity of the system increases for different network
of connected habitats and thus detailed study is involved with various
coupling aspects to hold this mechanisms of appearance and
disappearance of oscillations.

\begin{acknowledgments}

T.B. acknowledges the financial support from SERB, Department of
Science and Technology (DST), Govt. of India [grant:
  SB/FTP/PS-005/2013].  P.S.D.  acknowledges financial support from
SERB, Department of Science and Technology (DST), Govt. of India
[grant: YSS/2014/000057].

\end{acknowledgments}

\bibliography{tbref}

\end{document}